# Solutions of the Dirac equation in one-dimensional variable width potential well

Shan Qiuyu[†]

(college of science, Yanshan University, Qinhuangdao 066000, China)

Abstract

The Fermi acceleration mechanism is a significant source of cosmic rays. When the width of a potential well changes over time, the velocity of particles within the well also changes. For quantum systems, such dynamics should be described by the Schrödinger, Klein-Gordon, and Dirac equations. Previous studies have solved the Schrödinger and Klein-Gordon equations under these conditions, but no research has addressed the Dirac equation for spin-$\frac{1}{2}\hbar$ particles like electrons. This paper investigates the solutions of the Dirac equation in a dynamically varying potential well and demonstrates that Dirac particles can exhibit complex-valued momentum states via the Fermi acceleration mechanism, enabling Tachyon-like states preparation.

† Corresponding author.E-mail: shanqiuyu@outlook.com

## 1 Introduction

Solving the Schrödinger equation in a one-dimensional infinite potential well is a fundamental problem in quantum mechanics. For adiabatic systems, solutions can be obtained via separation of variables, representing superpositions of standing waves. However, when one wall of the potential well moves, the system becomes non-adiabatic, and the Hamiltonian becomes time-dependent. Fermi [1] and Ulam [2] revealed that particles in a contracting potential well (with one fixed wall and one moving wall) experience velocity amplification—the Fermi acceleration mechanism.

This phenomenon explains high-energy charged particle generation in astrophysical environments, such as bow shocks [3].

Quantum systems with moving potential walls are critical for studying Fermi acceleration. For non-relativistic particles (velocities $v \ll c$), the Schrödinger equation suffices. Doescher and Rice [4] derived exact solutions for such systems under static and uniformly moving walls. However, time-dependent Hamiltonians complicate exact solutions, necessitating approximations like the adiabaticor sudden approximation [5] . D.N.Pinder [6] explored these approximations for expanding/contracting wells, noting limitations in applying the sudden approximation to contraction cases. Recent studies also address pseudo-Hermitian Hamiltonians [7] and applications in atom optics [8].

Relativistic particles ($v \approx c$) require the Klein-Gordon or *Dirac equations*. While solutions for spin-0 particles (via the Klein-Gordon equation) exist [9–11], spin-$\frac{1}{2}\hbar$ particles (e.g., electrons) governed by the *Dirac equation* remain unexplored.

The *Dirac equation* in natural units ($\hbar = c = 1$)is:

$$i\frac{\partial\Psi}{\partial t} = -i\left(\alpha^1\frac{\partial\Psi}{\partial x^1} + \alpha^2\frac{\partial\Psi}{\partial x^2} + \alpha^3\frac{\partial\Psi}{\partial x^3}\right) + \beta m\Psi \tag{1}$$

where $\alpha^1$, $\alpha^2$, $\alpha^3$ and $\beta$ are Dirac matrices. Assuming spin-up particles which moves along the z-axis and using the Dirac-Pauli representation [12-13] , the *Dirac equation* should be written as

$$i\frac{\partial\Psi}{\partial t} = -i\alpha^z\frac{\partial\Psi}{\partial z} + \beta m\Psi \tag{2}$$

# 2 Solution of Dirac equation in one-dimensional motion well

### 2.1. Potential Well and Dirac Equation Setup

Consider a finite potential well with a moving right boundary:

$$U(z,t) = \begin{cases} U_0 & z \le 0 \\ 0 & 0 < z < vt \\ U_0 & z \ge vt \end{cases} \tag{3}$$

Where U0 is the potential height, v is the speed of the change of the coordinate of the well wall, the left side of the potential well wall is located at the origin of the z axis, the right side of the potential well wall moves with the speed v, and the width of the potential well changes accordingly.

In natural units($\hbar = c = 1$), the *Dirac equation* becomes:

$$i\frac{\partial\Psi(z,t)}{\partial t} = -i\alpha^z\frac{\partial\Psi(z,t)}{\partial z} + \beta m\Psi(z,t) + U(z,t)\Psi(z,t) \tag{4}$$

The Dirac Pauli representation requires that

$$\alpha^z = \begin{bmatrix} 0 & 0 & 1 & 0 \\ 0 & 0 & 0 & -1 \\ 1 & 0 & 0 & 0 \\ 0 & -1 & 0 & 0 \end{bmatrix}, \quad \beta = \begin{bmatrix} 1 & 0 & 0 & 0 \\ 0 & 1 & 0 & 0 \\ 0 & 0 & -1 & 0 \\ 0 & 0 & 0 & -1 \end{bmatrix}$$

the equation reduces to a coupled system for the spinor component $\Psi_0(z,t)$ and $\Psi_2(z,t)$.

$$\Psi(z,t)=\begin{bmatrix}\Psi_0(z,t)\\0\\\Psi_2(z,t)\\0\end{bmatrix} \tag{5}$$

The *Dirac equation* can be written in the form of the following equations:

$$\begin{cases} i\frac{\partial\Psi_0(z,t)}{\partial t}=-i\frac{\partial\Psi_2(z,t)}{\partial z}+(m+\mathrm{U(z,t)})\Psi_0(z,t)\\ i\frac{\partial\Psi_2(z,t)}{\partial t}=-i\frac{\partial\Psi_0(z,t)}{\partial z}+(-m+\mathrm{U(z,t)})\Psi_2(z,t)\end{cases} \tag{6}$$

If we set $A(z,t)=\frac{1}{2}[\Psi_0(z,t)+\Psi_2(z,t)]$, $B(z,t)=\frac{1}{2}[\Psi_0(z,t)-\Psi_2(z,t)]$, then the equation can be written as

$$\begin{cases} i\frac{\partial A(z,t)}{\partial t}=-i\frac{\partial A(z,t)}{\partial z}+mB(z,t)+\mathrm{U(z,t)}A(z,t)\\ i\frac{\partial B(z,t)}{\partial t}=i\frac{\partial B(z,t)}{\partial z}+mA(z,t)+\mathrm{U(z,t)}B(z,t)\end{cases} \tag{7}$$

And the boundary condition requires that $A(z,t)$ and $B(z,t)$are continuous.

## 2.2. Solutions in Different Regions

**Region I ( $\mathbf{z \leq 0}$):**

the above equations can be written as

$$\begin{cases} i\frac{\partial A(z,t)}{\partial t}=-i\frac{\partial A(z,t)}{\partial z}+mB(z,t)+\mathrm{U_0}A(z,t)\\ i\frac{\partial B(z,t)}{\partial t}=i\frac{\partial B(z,t)}{\partial z}+mA(z,t)+\mathrm{U_0}B(z,t)\end{cases} \tag{8}$$

solutions take the form:

$$A(z,t)=\sum_E[c_{1(\mathrm{E})}e^{-iEt-i\mathrm{k_1z}}] \tag{9}$$

$$B(z,t)=\frac{1}{m}\sum_E[c_{1(\mathrm{E})}(\mathrm{k_1}+E-\mathrm{U_0})e^{-iEt-i\mathrm{k_1z}}] \tag{10}$$

Where $\mathrm{k_1}=\sqrt{(\mathrm{E}-\mathrm{U_0})^2-\mathrm{m}^2}$, $c_{1(\mathrm{E})}$ is a constant related to E.

**Region II ($\mathbf{0 < z < vt}$):**

The equations are

$$\begin{cases} i\frac{\partial A(z,t)}{\partial t}=-i\frac{\partial A(z,t)}{\partial z}+mB(z,t)\\ i\frac{\partial B(z,t)}{\partial t}=i\frac{\partial B(z,t)}{\partial z}+mA(z,t)\end{cases} \tag{11}$$

We can get the equation of $A(z,t)$:

$$\frac{\partial^2A(z,t)}{\partial t^2}-\frac{\partial^2A(z,t)}{\partial z^2}+m^2A(z,t)=0 \tag{12}$$

By transforming coordinates[9–11] to $x=\sqrt{\frac{t+z}{t-z}}$ and $y=\sqrt{t^2-z^2}$, the equations decouple into Bessel functions.

$$-y^2\frac{\partial^2A(x,y)}{\partial y^2}+x^2\frac{\partial^2A(x,y)}{\partial x^2}-y\frac{\partial A(x,y)}{\partial y}+x\frac{\partial A(x,y)}{\partial x}-m^2y^2A(x,y)=0 \tag{13}$$

We can set that $A(x,y)=f(x)g(y)$:

$$\frac{x^2}{f(x)}\frac{d^2f(x)}{dx^2}+\frac{x}{f(x)}\frac{df(x)}{dx}=\frac{y^2}{g(y)}\frac{d^2g(y)}{dy^2}+\frac{y}{g(y)}\frac{dg(y)}{dy}-m^2g(y)=a \tag{14}$$

Solution:

$$A(x,y) = \sum_a (c_{2(a)}x^{\sqrt{a}} + c_{3(a)}x^{-\sqrt{a}})[c_{4(a)}J_{\sqrt{a}}(my) + c_5 Y_{\sqrt{a}}(my)] \quad (15)$$

$$B(x,y) = \sum_a i(c_{2(a)}x^{\sqrt{a}-1} + c_{3(a)}x^{-\sqrt{a}-1})[c_{4(a)}J_{\sqrt{a}-1}(my) + c_5 Y_{\sqrt{a}-1}(my)] \quad (16)$$

and $J_{\sqrt{a}}(my)$ and $Y_{\sqrt{a}}(my)$are the Bessel function.

In order to avoid the divergence of $Y_{\sqrt{a}}(my)$ at $y = 0$ let's set $c_5 = 0$.

$$A(x,y) = \sum_a (c_{2(a)}x^{\sqrt{a}} + c_{3(a)}x^{-\sqrt{a}})c_{4(a)}J_{\sqrt{a}}(my) \quad (17)$$

$$B(x,y) = \sum_a i(c_{2(a)}x^{\sqrt{a}-1} + c_{3(a)}x^{-\sqrt{a}-1})c_{4(a)}J_{\sqrt{a}-1}(my) \quad (18)$$

**Region III (z ≥ vt):**

Solutions involve exponential functions, derived via coordinate substitutions $p = vt - z$ and $q = t$.

$$A(p,q) = \sum_{E'} c_{6(E')}e^{ik(vq-p)-iE'q} + c_{7(E')}e^{-ik(vq-p)-iE'q} \quad (19)$$

$$B(z,t) = \frac{1}{m}\sum_{E'} c_{6(E')}(E' - k - \mathrm{U}_0)e^{ik(vq-p)-iE'q} + c_{7(E')}(k + E' - \mathrm{U}_0)e^{-ik(vq-p)-iE'q}$$

(20)

Where $k = \sqrt{(E' - \mathrm{U}_0)^2 - \mathrm{m}^2}$.

### 2.3. Boundary Matching

At $z = 0$, continuity of $A$ and $B$ leads to transcendental equations for coefficients

$$\sum_E c_{1(\mathrm{E})}e^{-iEt} = \sum_a (c_{2(a)} + c_{3(a)})c_{4(a)}J_{\sqrt{a}}(my) \quad (21)$$

$$\frac{1}{m}\sum_E c_{1(\mathrm{E})}(\mathrm{k}_1 + E - \mathrm{U}_0)e^{-iEt} = \sum_a i(c_{2(a)} + c_{3(a)})c_{4(a)}J_{\sqrt{a}-1}(mt) \quad (22)$$

Expanding Bessel functions in Taylor series and matching terms at $t = 0$ reveals that only integer-order solutions satisfy the boundary conditions (21):

$$\sum_E c_{1(\mathrm{E})}e^{-iEt} = \sum_{l=0}^{\infty}[\sum_E c_{1(\mathrm{E})}\frac{(iE)^l}{l!}]\mathrm{t}^{\mathrm{l}} \quad (23)$$

$$J_{\sqrt{a}}(mt) = \sum_{k=0}^{\infty}\frac{(-1)^k}{k!\Gamma(k+1+\sqrt{a})}(\frac{mt}{2})^{2k+\sqrt{a}} \quad (24)$$

Since $l$ is a natural number, we infer that $a = n^2,\ n = 0,1,2,3\ldots$(It is obvious but a more detailed mathematical proof is shown in Appendix Ⅱ)

The difference between the results obtained through the Klein Gordon equation[9] and here is due to the Klein tunneling[12] of Dirac particles.Where Dirac particles penetrate high barriers via particle-antiparticle conversion, necessitating continuous boundary conditions rather than vanishing wavefunctions.

Consider the zero-degree term in the expansion of the boundary condition (21)

$$\sum_E c_{1(\mathrm{E})} = (c_{2(0)} + c_{3(0)})c_{4(0)} \quad (25)$$

If $\sum_E c_{1(\mathrm{E})} \neq 0$, the term $a = 0$ must be existing.

At large $t$, the solutions at $z = 0$ asymptotically approach oscillatory modes dominated by $\sum_a (c_{2(a)} + c_{3(a)})c_{4(a)}\sqrt{\frac{2}{\pi mt}} cos(mt - \frac{\pi\sqrt{a}}{2} - \frac{\pi}{4})$, the solutions should be a superposition state of positive and negative energy solutions.

The solution of the *Dirac equation* in a potential well of the shape (3) is:

$$\begin{cases} \Psi_0(z,t) = A(z,t) + B(z,t) \\ \Psi_2(z,t) = A(z,t) - B(z,t) \end{cases} \tag{26}$$

Fig. 1 show us the shape of a numerical solution at $\mathrm{t} = 1$and $\mathrm{t} = 5$, where $\mathrm{m} = 1$ (an electron mass in natural units) , $\mathrm{U_0} = 1$, $\mathrm{v} = 0.6$, It's the simplest free particle solution (unnormalized) when $\mathrm{t} \leq 0$: $A(z,t) = B(z,t) = e^{-2it}$

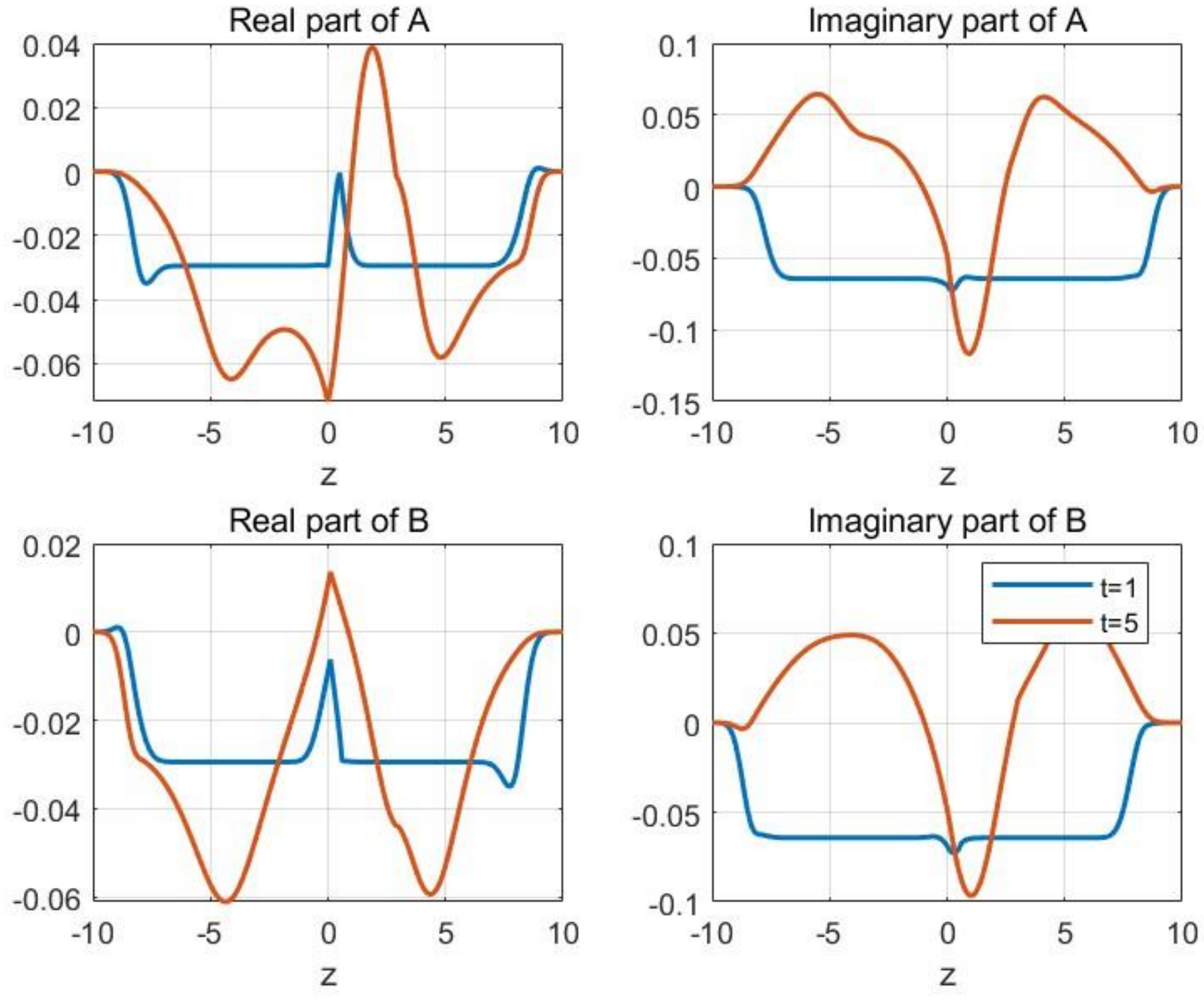

Fig. 1 numerical solution at $\mathrm{t} = 1$and $\mathrm{t} = 5$

Calculate the average momentum of this quantum state in Region II, which is approximately -0.000369 + -0.000119i at t=1 and -0.005637 + -0.001718i at t=5. This indicates that Dirac particles can have imaginary momentum in such potential wells.

## 3 complex-valued momentum and Tachyon-like states

### 3.1.Complex Momentum Analysis

Assume the solution at time t is:

$$\begin{cases} \Psi_0(z,t) = J_0(m\sqrt{t^2 - z^2}) - i(\sqrt{\frac{t-z}{t+z}})J_1(m\sqrt{t^2 - z^2}) \\ \Psi_2(z,t) = J_0(m\sqrt{t^2 - z^2}) + i(\sqrt{\frac{t-z}{t+z}})J_1(m\sqrt{t^2 - z^2}) \end{cases} \tag{27}$$

As we have seen in section 2.3, If $\sum_E c_{1(E)} \neq 0$, such term is inevitable.

Let's set that $F(z) = J_0(m\sqrt{t^2 - z^2})$ and $G(z) = (\sqrt{\frac{t-z}{t+z}})J_1(m\sqrt{t^2 - z^2})$. For $v < c$, $F(z)$ and $G(z)$ are real functions. The momentum expectation at time t is:

$$< \hat{p} >= i\int_0^{vt} [F(z) + iG(z)][\frac{\partial F(z)}{\partial z} - i\frac{\partial G(z)}{\partial z}] + [F(z) - iG(z)][\frac{\partial F(z)}{\partial z} + i\frac{\partial G(z)}{\partial z}]dz \tag{28}$$

Simplifying yields:

$$< \hat{p} >= i\int_0^{vt} 2[F(z)\frac{\partial F(z)}{\partial z} + G(z)\frac{\partial G(z)}{\partial z}]dz \tag{29}$$

$< \hat{p} >$ is time-dependent and becomes purely imaginary, even for $v < c$.

This implies complex-valued momentum states in the zero-potential region, distinct from tunneling.

The relationship between momentum p and velocity u is:

$$p = \frac{mu}{\sqrt{1-u^2}} \tag{30}$$

Imaginary momentum implies $u > c$. However, these solutions are unstable due to the non-constant modulus of Bessel functions and represent superpositions of positive- and negative-energy states.

In addition, we set $v = 0.6$ and $m = 1$ to investigate the variation of the imaginary part of momentum expectation with time. The results are shown in Fig. 2, which shows that modulus of momentum oscillates and decreases. When t is small, there is a rapid decrease in the momentum, which reaches its minimum value near t=0.28.

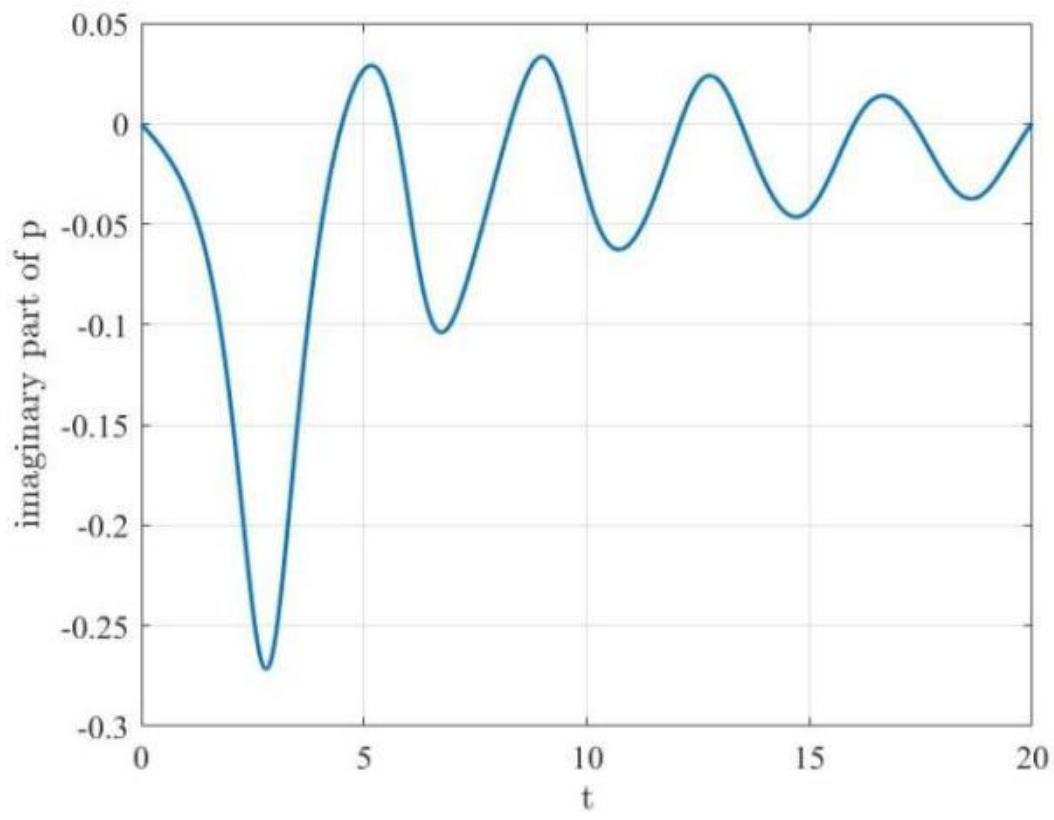

Fig. 2 imaginary part of momentum expectation varies over time, there is a rapid increase in modulus at first and then followed by a decrease in oscillation over time

### 3.2.Wave Packet Dynamics

Examining the motion of the wave packet center ($\frac{\partial\Psi^*\Psi}{\partial z}=0$) of quantum state (27), the velocity of the center (see Fig. 3) , u , exceeds the speed of light ( $c=1$ ), demonstrating Tachyon-like behavior.

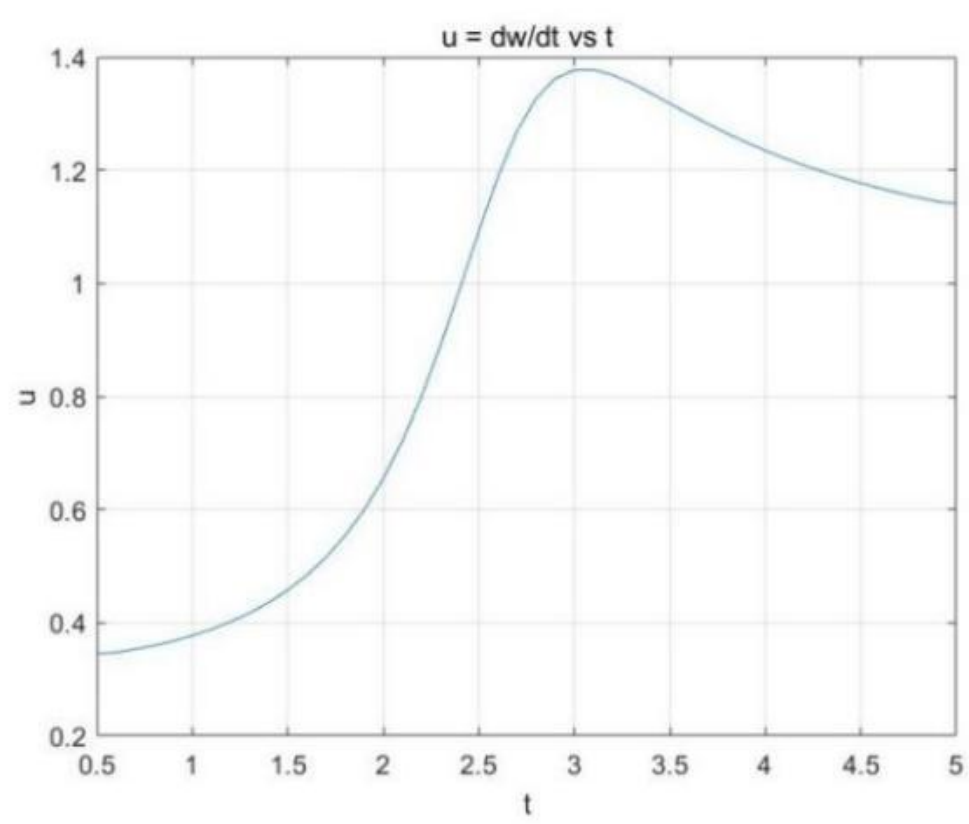

Fig. 3 relationship between wave packet center velocity and time

We can measure the position and velocity of the center of the wave packet, and if a quantum state with imaginary momentum is generated, u should be faster than light.

The Lagrangian of the free Dirac field [14] is

$$\mathcal{L}=\overline{\Psi}(i\gamma^{\mu}\partial_{\mu}-m)\Psi \tag{31}$$

and the energy-momentum tensor:

$$T^{\mu}_{\nu}=\frac{\partial\mathcal{L}}{\partial(\partial_{\mu}\Psi)}\partial_{\nu}\Psi-\mathcal{L}\delta^{\mu}_{\nu} \tag{32}$$

or we can write it as

$$T^{\mu\nu}=\overline{\Psi}i\eta^{\rho\nu}\gamma^{\mu}\frac{\partial\Psi}{\partial x^{\rho}}-\mathcal{L}\eta^{\mu\nu} \tag{33}$$

Where $\eta^{\mu\nu}$ is the Minkowski metric, $\gamma^{\mu}$ is Dirac matrix, $\overline{\Psi}=\Psi^*\gamma^0$, using Einstein's summation convention. We can use (33) to calculate the energy density $T^{00}$ and energy flow density $T^{03}$ , note that $\Psi_0(z,t)$ and $\Psi_2(z,t)$ are conjugate complex in (27), and a simple calculation can yield:

$$T^{03}=i\frac{\partial}{\partial z}({\Psi_0}^*\Psi_0) \tag{34}$$

$$T^{00}=i\Psi_0\frac{\partial\Psi_0}{\partial z}+i\Psi_2\frac{\partial\Psi_2}{\partial z} \tag{35}$$

The energy flow velocity of quantum state (27) is :

$$v_{ener}=\frac{T^{03}}{T^{00}} \tag{36}$$

Fig.4 shows the maximum energy flow velocity $v_{ener}$ at $t$, since $T^{00}$ can be close to 0, $v_{ener}$ could be much higher than the speed of light. It can be observed experimentally, suggesting that the imaginary momentum here correspond to tachyon-like motion and not to a mathematical artifact.

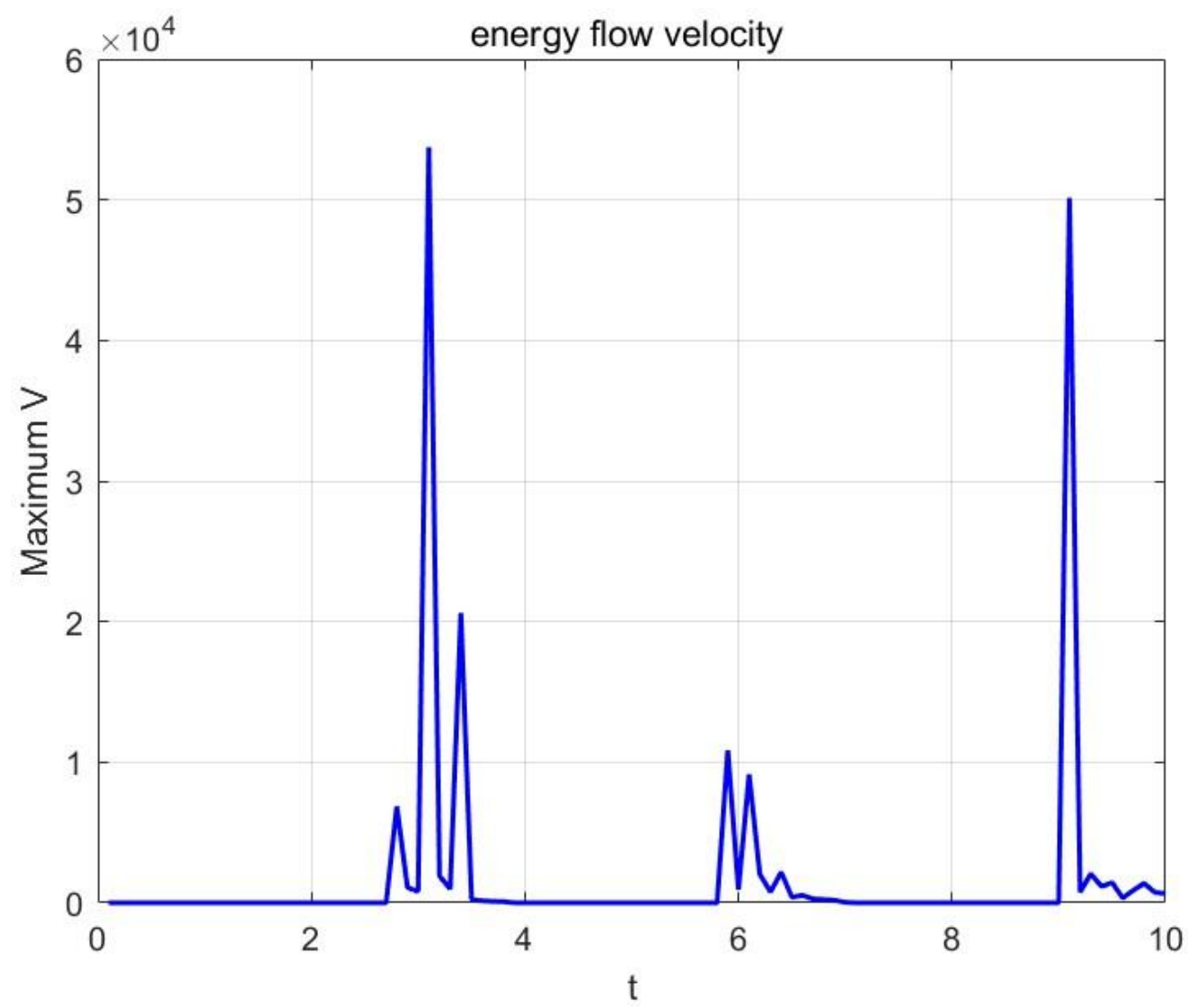

Fig. 4 The change of maximum energy flow velocity with time

Set time $t = 2.7$, $v = 0.6$ and $m = 1$, the distribution of energy flow velocity in Region II is shown in Fig.5.

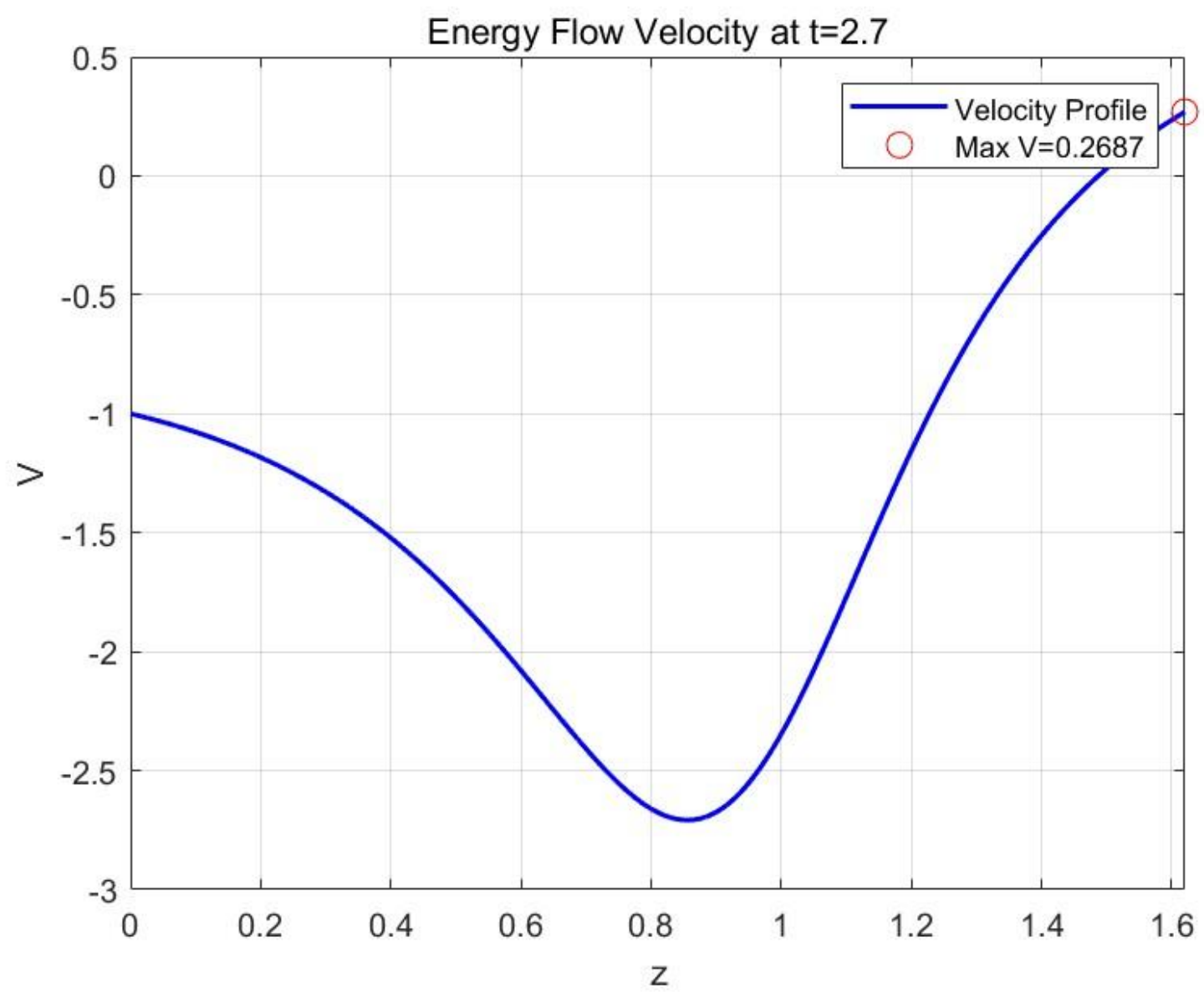

Fig. 5 The spatial distribution of energy flow velocity at t=2.7 , its average value in Region II is about -1.4

Consider the average energy flow velocity of the time interval $[0, T]$:

$$v_{avag} = \frac{1}{T}\int_0^T \left(\frac{1}{0.6t}\int_0^{0.6t} v_{ener}\, dz\right)dt \tag{37}$$

The value for $T = 3$ is about -2.8.

The probability density and the probability flux density in the z-direction are

$$\begin{cases} \rho = \Psi^*\Psi = 2\Psi_0\Psi_2 \\ j_z = \Psi^*\alpha^z\Psi = \Psi_0\Psi_0 + \Psi_2\Psi_2 \end{cases} \tag{38}$$

The probability flow velocity is

$$v_{prob} = \frac{j_z}{\rho} \tag{39}$$

$v_{prob}$ is also a complex, but Fig.6 shows the relationship between the real part of

$v_{prob}$ and time, it can be seen that the real part can be greater than the speed of light.

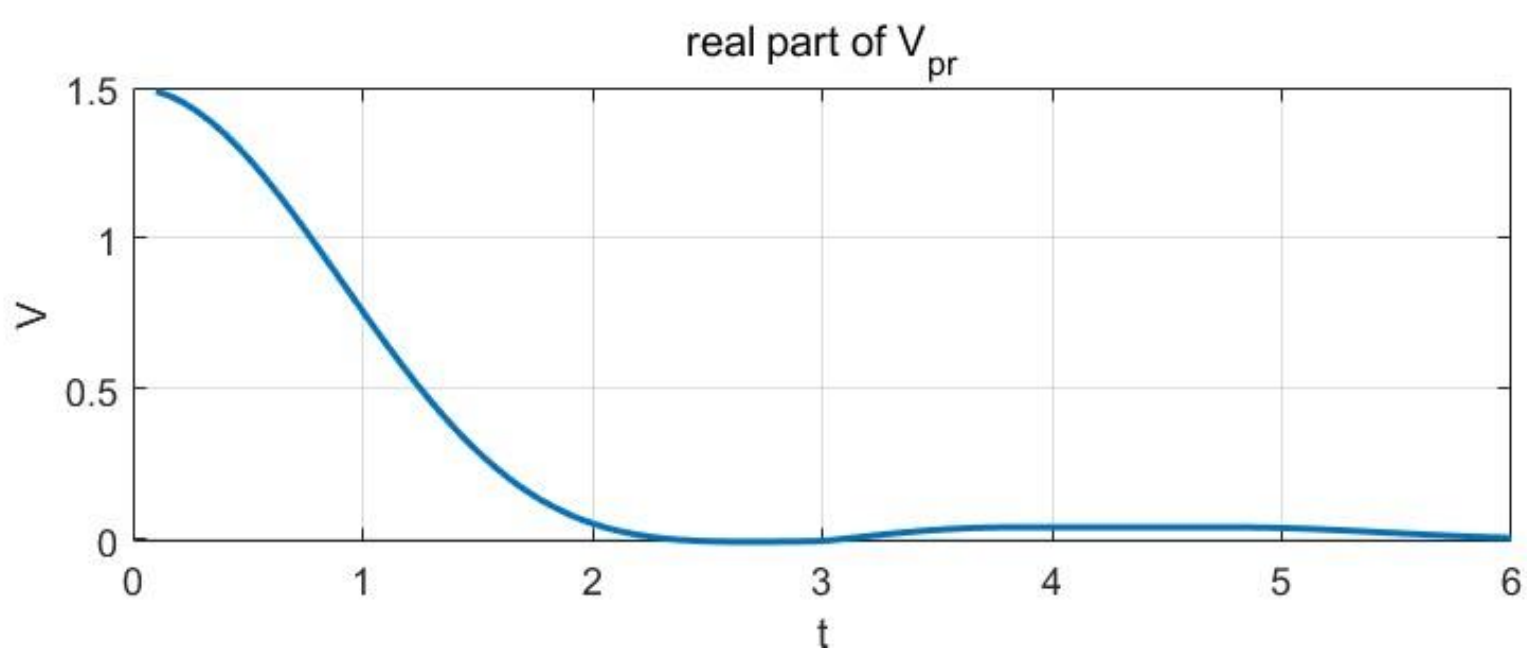

Fig. 6 real part of the probability flow velocity varies with time, The well wall exhibited an initial velocity peak at the onset of motion, followed by a rapid deceleration.

# 4 conclusion

We derive analytical solutions for the *Dirac equation* in a variable-width potential well. Key findings include:

**1. Relativistic Superposition:** The solutions in $0 < z < vt$ are superpositions of Bessel functions with quantized orders, reflecting both positive and negative energy states.

**2. Klein Tunneling Impact:** Unlike the Schrödinger or Klein-Gordon cases, the Dirac solutions require continuous boundary conditions due to Klein tunneling, preventing infinite confinement.

**3. Numerical Necessity:** Exact parameter matching demands numerical methods, but analytical forms provide insight into the system's spectral and dynamical properties.

**4. complex-valued momentum and Tachyon-like states:** Dirac particles in variable-width potential wells exhibit complex-valued momentum and the energy flow velocity could be faster than light, enabling superluminal quantum states without requiring $v > c$.

**5. Limitations and possible follow-up research:** Relativity prohibits tachyon, and we are also unsure whether the *Dirac equation* can describe tachyon-like states. It means that we may need to modify Relativity or *Dirac equation* to describe tachyon (for example, using other particles rather than light to calibrate the simultaneity in Relativity), or use the modified theory to make particle momentum a real-valued. Furthermore, the one-dimensional finite-depth potential well serves as a simplified theoretical framework, complex-valued momentum states may exhibit widespread prevalence in time-dependent Dirac systems, this phenomenon necessitates more rigorous theoretical elucidation coupled with systematic experimental verification.

This work bridges a critical gap in relativistic quantum mechanics and offers a foundation for studying fermionic systems in astrophysical and laboratory environments.

# Acknowledgments

Thanks DeepSeek AI for assistance in literature curation, numerical code verification, and language polishing. DeepSeek AI did not participate in theoretical derivations or result interpretation.

## Appendix Ⅰ Superluminal potential wall velocity analysis.

As an interesting case, We investigate the scenario where the potential well's wall velocity faster that light.

Consider uniformly distributed identical charged particles on the XOY plane, where each particle corresponds to a coordinate $(x, y)$. Through a method such as the stimulated Raman process, charged particles within the region $0 < x < x0$ are displaced from $(x, y)$ to $(x, y + w)$.

$w = w0$ if $y > 0$ and $w = -w0$ if $y < 0$, where $w0$ is a positive real number. This configuration establishes a potential well in the region $0 < x < x0$ , $-w0 < y < w0$.

$$U(x, y, t) = \begin{cases} 0 & 0 < x < x0\,,\ -w0 < y < w0 \\ U_0 & others \end{cases} \tag{A1}$$

The potential well can be described using a global parameter $w(x, y)$ . The potential well walls can be made to "move" in an apparent manner. For example, consider the following piecewise function:

$$w(x, y) = \begin{cases} w0 & x > 0 \\ 0 & 0 < x < vt \\ w0 & x > vt \end{cases} \tag{A2}$$

Here, $v$ represents the apparent velocity of the potential well wall. Crucially, this "motion" corresponds to an instantaneous reconstruction process—either the collapse or formation of the well walls—rather than true physical movement. Analogous to dominoes falling under controlled triggering (rather than mechanical pushing by preceding dominoes), this apparent velocity $v$ can exceed the speed of light without violating relativistic principles.

Numerical solutions for *v*=2 are provided by Fig. 7. (other parameters consistent with Fig. 1).

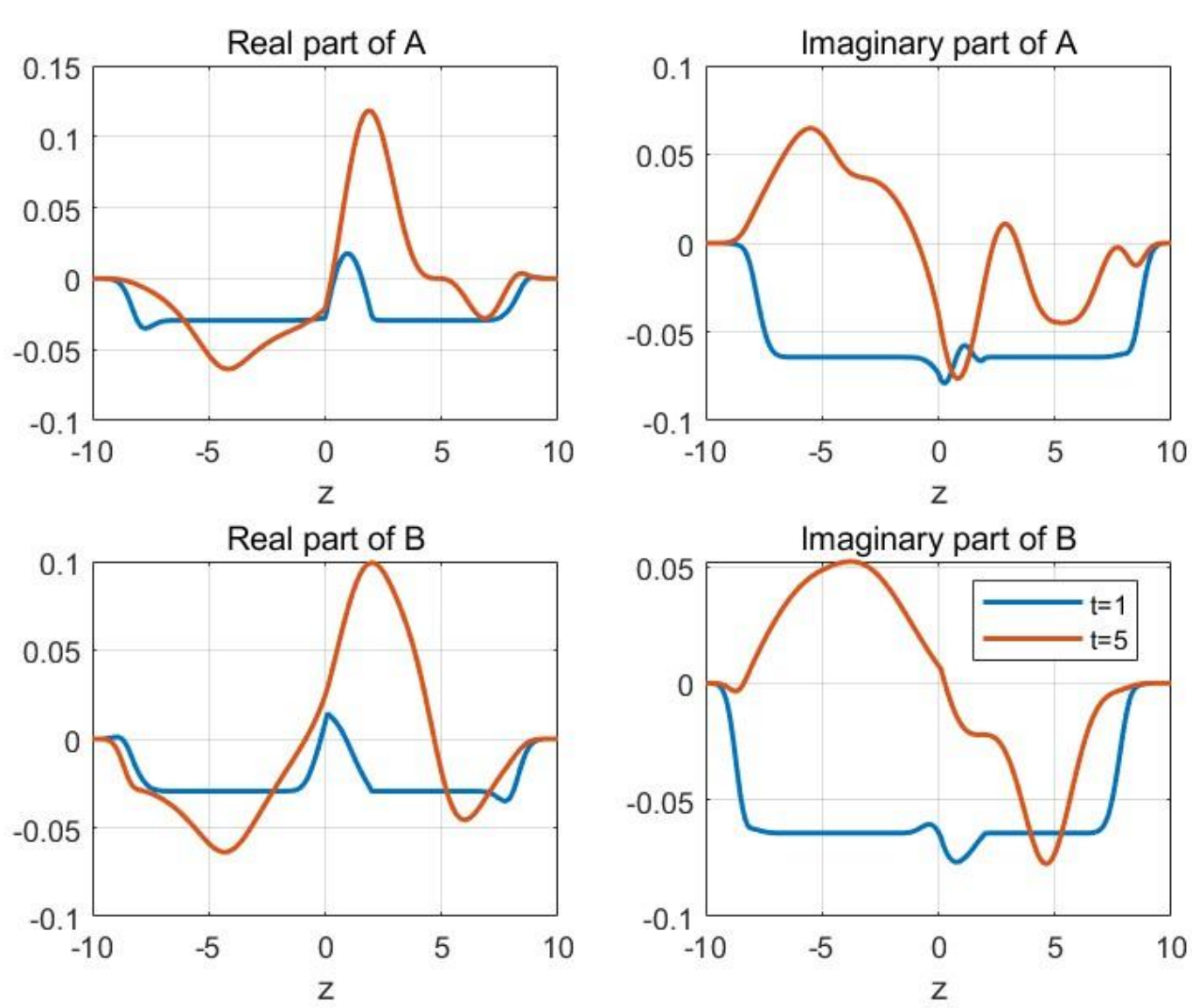

Fig. 7 Numerical solutions for *v*=2, compared with *v*=0.6, the shape is similar at t=1, but there is a significant difference at t=5

In this case, the solution obtained in Ref. [9] exhibits the unavoidable divergence. But we can also construct another kind of potential well:

$$w(x,y)=\begin{cases} w0 & x>vt \\ 0 & vt<x<vt+L \\ w0 & x>vt+L \end{cases} \tag{A3}$$

It appears to be a static well in another frame of reference, but if we consider it as a one-dimensional infinite well, the solution to the Klein-Gordon equation should be

$$\Psi = Asin[\frac{n\pi}{L}(x-vt)]e^{-i\sqrt{\frac{n^2\pi^2(1-v^2)}{L^2}+m^2}\frac{t-vx}{\sqrt{1-v^2}}} \tag{A4}$$

If $v>c$, its momentum expected value will also be complex. Of course, it is uncertain whether the model of a one-dimensional infinite well with $v>c$ can be used, so it is provided here only for reference.

## Appendix Ⅱ

In the original, (23) and (24) are:

$$\sum_E c_{1(\mathrm{E})}e^{-iEt} = \sum_{l=0}^{\infty}[\sum_E c_{1(\mathrm{E})}\frac{(iE)^l}{l!}]\mathrm{t}^{\mathrm{l}} \tag{23}$$

$$J_{\sqrt{a}}(mt)=\sum_{k=0}^{\infty}\frac{(-1)^k}{k!\Gamma(k+1+\sqrt{a})}(\frac{mt}{2})^{2k+\sqrt{a}} \tag{24}$$

Boundary conditions (21) require that the above two expressions be equal:

$$\sum_{l=0}^{\infty}[\sum_E c_{1(E)}\frac{(iE)^l}{l!}]t^l=\sum_{k=0}^{\infty}\frac{(-1)^k}{k!\Gamma(k+1+\sqrt{a})}(\frac{mt}{2})^{2k+\sqrt{a}} \tag{A5}$$

It means that the powers of $t$ on both sides of the equal sign need to match exactly, i.e. for any $a$ there is at least one array$(k,l)$ such that $\sqrt{a}=l-2k$.

Since both $k$ and $l$ are natural numbers, $\sqrt{a}$ must be natural.

In addition, as $t$ approaches 0 the Bessel function approximates

$$J_{\sqrt{a}}(mt)\approx\frac{1}{\Gamma(1+\sqrt{a})}(\frac{mt}{2})^{\sqrt{a}} \tag{A6}$$

Its N-th derivative is

$$\frac{d^n}{dt^n}J_{\sqrt{a}}(mt)\approx\frac{m^{\sqrt{a}}\sqrt{a}(\sqrt{a}-1)(\sqrt{a}-2)\ldots(\sqrt{a}-n+1)}{2^{\sqrt{a}}\Gamma(1+\sqrt{a})}t^{\sqrt{a}-n} \tag{A7}$$

If $\sqrt{a}$ is a non integer, then if $\mathrm{n}>\sqrt{a}$, At $t=0$, the n-th derivative of $J_{\sqrt{a}}(mt)$ will diverge, which does not meet the requirement of infinite differentiability of wave functions in physics and contradicts the form of infinite differentiability on the left side of the equation (A2).

Therefore, we can get that $a=n^2,\ n=0,1,2,3\ldots$

Of course, more constraints on $a$ need to match specific boundary conditions and initial conditions.

# Appendix Ⅲ (Numerical Calculation Code)

**Fig. 1(Including stability, error and convergence analysis)**

```
v = 0.6;
m = 1;
U0 = 1;
z_min = -10;
z_max = 10;
t_end = 6;
cfl = 0.1;
base_points = 201;
damp_width = 2.0;
sigma_max = 20.0;
dz_base = (z_max - z_min)/(base_points-1);
dt_base = cfl*dz_base;
z_base = linspace(z_min, z_max, base_points)';
[A_base, B_base] = DiracSolver(base_points, dz_base, dt_base, t_end, ...
z_min, z_max, v, m, U0, damp_width, sigma_max);
t_targets = [1, 5];
for i = 1:length(t_targets)
t_target = t_targets(i);
vt_val = v * t_target;
A_t = A_base(:,i);
B_t = B_base(:,i);
A_conj = conj(A_t);
B_conj = conj(B_t);
dz = dz_base;
dA_dz = gradient(A_t, dz);
dB_dz = gradient(B_t, dz);
integrand = A_conj .* dA_dz + B_conj .* dB_dz;
indices = (z_base >= 0) & (z_base <= vt_val);
p_avg = (1i/2) * trapz(z_base(indices), integrand(indices));
fprintf('t=%d, <p̂> = %.6f + %.6fi\n', t_target, real(p_avg), imag(p_avg));
end
figure;
t_save = [1, 5];
for i = 1:2
subplot(2,2,1)
plot(z_base, real(A_base(:,i)), 'LineWidth',1.5,
'DisplayName',['t=',num2str(t_save(i))]);
hold on; title('real A'); xlabel('z'); grid on;
subplot(2,2,2)
plot(z_base, imag(A_base(:,i)), 'LineWidth',1.5,
'DisplayName',['t=',num2str(t_save(i))]);
hold on; title('imag A'); xlabel('z'); grid on;
subplot(2,2,3)
```

```
plot(z_base, real(B_base(:,i)), 'LineWidth',1.5,
'DisplayName',['t=',num2str(t_save(i))]);
hold on; title('real B'); xlabel('z'); grid on;
subplot(2,2,4)
plot(z_base, imag(B_base(:,i)), 'LineWidth',1.5,
'DisplayName',['t=',num2str(t_save(i))]);
hold on; title('imag B'); xlabel('z'); grid on;
end
legend('show','Location','best');
ref_levels = 4;
[orders, errors] = runConvergenceTest(base_points, ref_levels, t_end, ...
z_min, z_max, v, m, U0, damp_width, sigma_max, cfl);
fprintf('convergence order (A): '); fprintf('%.2f ', orders.A); fprintf('\n');
fprintf('convergence order (B): '); fprintf('%.2f ', orders.B); fprintf('\n');
fprintf('L2 error (A): '); fprintf('%.2e ', errors.A); fprintf('\n');
fprintf('L2 error (B): '); fprintf('%.2e ', errors.B); fprintf('\n');
function [orders, errors] = runConvergenceTest(N_base, levels, t_end, ...
z_min, z_max, v, m, U0, damp_width, sigma_max, cfl)
N_ref = N_base * 2^levels;
dz_ref = (z_max - z_min)/(N_ref-1);
dt_ref = cfl*dz_ref;
[A_ref, B_ref] = DiracSolver(N_ref, dz_ref, dt_ref, t_end, ...
z_min, z_max, v, m, U0, damp_width, sigma_max);
errors.A = zeros(1, levels);
errors.B = zeros(1, levels);
orders.A = zeros(1, levels-1);
orders.B = zeros(1, levels-1);
for lev = 1:levels
N = N_base * 2^(lev-1);
dz = (z_max - z_min)/(N-1);
dt = cfl*dz;
[A_sol, B_sol] = DiracSolver(N, dz, dt, t_end, ...
z_min, z_max, v, m, U0, damp_width, sigma_max);
z_ref = linspace(z_min, z_max, N_ref)';
A_interp = interp1(linspace(z_min,z_max,N)', A_sol(:,2), z_ref, 'spline');
B_interp = interp1(linspace(z_min,z_max,N)', B_sol(:,2), z_ref, 'spline');
errors.A(lev) = norm(A_interp - A_ref(:,2)) * sqrt(dz_ref);
errors.B(lev) = norm(B_interp - B_ref(:,2)) * sqrt(dz_ref);
end
for lev = 1:levels-1
orders.A(lev) = log(errors.A(lev)/errors.A(lev+1)) / log(2);
orders.B(lev) = log(errors.B(lev)/errors.B(lev+1)) / log(2);
end
end
function [A_save, B_save] = DiracSolver(Nz, dz, dt, t_total, z_min, z_max, ...
v, m, U0, damp_width, sigma_max)
```

```
Nt = ceil(t_total/dt);
dt = t_total/Nt;
z = linspace(z_min, z_max, Nz)';
damp = zeros(Nz,1);
for j = 1:Nz
if z(j) < z_min + damp_width
ratio = (z_min + damp_width - z(j))/damp_width;
damp(j) = sigma_max * ratio^2;
elseif z(j) > z_max - damp_width
ratio = (z(j) - (z_max - damp_width))/damp_width;
damp(j) = sigma_max * ratio^2;
end
end
norm_factor = sqrt(dz/(z_max - z_min));
A = ones(Nz,1) * norm_factor;
B = ones(Nz,1) * norm_factor;
save_steps = round([1,5]/dt) + 1;
A_save = zeros(Nz, 2);
B_save = zeros(Nz, 2);
save_idx = 1;
for n = 1:Nt
t = (n-1)*dt;
if ismember(n, save_steps)
A_save(:,save_idx) = A;
B_save(:,save_idx) = B;
save_idx = save_idx + 1;
end
[k1A, k1B] = waveRHS(A, B, dz, m, U0, v, t, z_min, z_max, damp);
A1 = A + 0.5*dt*k1A;
B1 = B + 0.5*dt*k1B;
[k2A, k2B] = waveRHS(A1, B1, dz, m, U0, v, t+0.5*dt, z_min, z_max, damp);
A2 = A + 0.5*dt*k2A;
B2 = B + 0.5*dt*k2B;
[k3A, k3B] = waveRHS(A2, B2, dz, m, U0, v, t+0.5*dt, z_min, z_max, damp);
A3 = A + dt*k3A;
B3 = B + dt*k3B;
[k4A, k4B] = waveRHS(A3, B3, dz, m, U0, v, t+dt, z_min, z_max, damp);
A = A + dt/6*(k1A + 2*k2A + 2*k3A + k4A);
B = B + dt/6*(k1B + 2*k2B + 2*k3B + k4B);
end
end
function [dAdt, dBdt] = waveRHS(A, B, dz, m, U0, v, t, z_min, z_max, damp)
Nz = length(A);
z = linspace(z_min, z_max, Nz)';
vt = v*t;
U = U0 * ~((z > 0) & (z < vt));
```

```
dAdt = zeros(Nz,1);
dBdt = zeros(Nz,1);
for j = 2:Nz
dA_dz = (A(j) - A(j-1))/dz;
dAdt(j) = -dA_dz - 1i*(m*B(j) + U(j)*A(j)) - damp(j)*A(j);
end
dAdt(1) = -(A(2)-A(1))/dz - 1i*(m*B(1)+U(1)*A(1)) - damp(1)*A(1);

for j = 1:Nz-1
dB_dz = (B(j+1) - B(j))/dz;
dBdt(j) = dB_dz - 1i*(m*A(j) + U(j)*B(j)) - damp(j)*B(j);
end
dBdt(Nz) = (B(Nz)-B(Nz-1))/dz - 1i*(m*A(Nz)+U(Nz)*B(Nz)) - damp(Nz)*B(Nz);
End
```

## Fig. 2

```
v = 0.6;
m = 1;
t_start = 0.01;
t_end = 20;
num_points = 1000;
t_values = linspace(t_start, t_end, num_points);
p_expectations = zeros(1, num_points);
h = 1e-6;
for i = 1:length(t_values)
t = t_values(i);
vt = v * t;
f = @(z) besselj(0, m*sqrt(t^2 - z.^2));
g = @(z) sqrt((t - z)./(t + z)) .* besselj(1, m*sqrt(t^2 - z.^2));
integrand_A = @(z) f(z).^2 + g(z).^2;
A = 2 * integral(integrand_A, 0, vt, 'RelTol', 1e-6, 'AbsTol', 1e-9);
df = @(z) (f(z + h) - f(z - h)) / (2*h);
dg = @(z) (g(z + h) - g(z - h)) / (2*h);
integrand_p = @(z) 2 * (f(z) .* arrayfun(df, z) + g(z) .* arrayfun(dg, z));
integral_p_val = integral(integrand_p, 0, vt, 'ArrayValued', true, 'RelTol',
1e-6);
p_expectation = 1i * integral_p_val / A;
p_expectations(i) =imag(p_expectation) ;
end
figure;
plot(t_values, p_expectations, 'LineWidth', 1.5);
xlabel('t', 'Interpreter', 'latex');
ylabel('imaginary part of p', 'Interpreter', 'latex');
grid on;
box on;
```

```
set(gca, 'FontSize', 12, 'FontName', 'Times New Roman');
```

## Fig. 3

```
t_values = 0.5:0.1:5;
w_values = zeros(size(t_values));
options = optimset('Display', 'off', 'TolFun', 1e-6);
for i = 1:length(t_values)
t = t_values(i);
if i == 1
initial_guess = 0;
else
initial_guess = w_values(i-1);
end
z_sol = fsolve(@(z)dAdz(z, t), initial_guess, options);
w_values(i) = z_sol;
end
dt = t_values(2) - t_values(1);
u_values = gradient(w_values, dt);
if any(u_values > 1)
disp('u>1');
else
disp('u<=1');
end
figure;
plot(t_values, u_values);
xlabel('t');
ylabel('u');
title('u = dw/dt vs t');
grid on;
function F = dAdz(z, t)
s = sqrt(t^2 - z^2);
if isreal(s) && s ~= 0
J0 = besselj(0, s);
J1 = besselj(1, s);
J0_prime = -J1;
J1_prime = (besselj(0, s) - besselj(2, s)) / 2;
dsdz = -z / s;
term1 = 2 * J0 * J0_prime * dsdz;
term2_part1 = (-2*t) / (t + z)^2 * J1^2;
term2_part2 = ((t - z)/(t + z)) * 2 * J1 * J1_prime * dsdz;
term2 = term2_part1 + term2_part2;
F = 2 * (term1 + term2);
else
F = Inf;
end
End
```

## Fig. 4

```
t_values = 0.1:0.1:10;
max_V = zeros(size(t_values));
for i = 1:length(t_values)
t = t_values(i);
z_max = 0.6 * t;
dz = 0.001;
z = 0:dz:z_max;
if isempty(z)
z = [0, dz];
end
sqrt_arg = sqrt(t^2 - z.^2);
J0 = besselj(0, sqrt_arg);
J1 = besselj(1, sqrt_arg);
sqrt_factor = sqrt((t - z) ./ (t + z));
Psi0 = J0 - 1i * sqrt_factor .* J1;
Psi2 = J0 + 1i * sqrt_factor .* J1;
dPsi0_dz = gradient(Psi0, dz);
dPsi2_dz = gradient(Psi2, dz);
T00 = real(Psi0 .* dPsi0_dz + Psi2 .* dPsi2_dz);
T03 = real(Psi2 .* dPsi0_dz + Psi0 .* dPsi2_dz);
V = abs(T03 ./ T00);
V_valid = V(isfinite(V));
if ~isempty(V_valid)
max_V(i) = max(V_valid);
else
max_V(i) = NaN;
end
end
figure;
plot(t_values, max_V, 'b-', 'LineWidth', 1.5);
xlabel('t');
ylabel('Maximum V');
title('energy flow velocity ');
grid on;
```

## Fig. 5

```
t = 2.7;
z_max = 0.6 * t;
dz = 0.001;
z = 0:dz:z_max;
if isempty(z)
z = [0, dz];
end
```

```
sqrt_arg = sqrt(t^2 - z.^2);
J0 = besselj(0, sqrt_arg);
J1 = besselj(1, sqrt_arg);
sqrt_factor = sqrt((t - z)./(t + z));
Psi0 = J0 - 1i * sqrt_factor .* J1;
Psi2 = J0 + 1i * sqrt_factor .* J1;
dPsi0_dz = gradient(Psi0, dz);
dPsi2_dz = gradient(Psi2, dz);
T00 = real(Psi0 .* dPsi0_dz + Psi2 .* dPsi2_dz);
T03 = real(Psi2 .* dPsi0_dz + Psi0 .* dPsi2_dz);
V = T03 ./ T00;
valid_idx = isfinite(V);
z_valid = z(valid_idx);
V_valid = V(valid_idx);
integral_result = (1/1.62) * trapz(z_valid, V_valid);
fprintf(' %.6f\n', integral_result);
figure;
plot(z_valid, V_valid, 'b-', 'LineWidth', 1.5);
xlabel('z');
ylabel('V');
title(['Energy Flow Velocity at t=', num2str(t)]);
grid on;
[V_max, max_idx] = max(V_valid);
hold on;
plot(z_valid(max_idx), V_max, 'ro', 'MarkerSize', 8);
legend('Velocity Profile', ['Max V=', num2str(V_max, '%.4f')]);
xlim([0 z_max]);
```

## Fig. 6

```
syms z t real;
m = 1;
A = besselj(0, m*sqrt(t^2 - z^2));
B = -1i*sqrt((t - z)/(t + z)) * besselj(1, m*sqrt(t^2 - z^2));
v=real((((A+B)^2)+((A-B)^2)/2)*(A+B)*(A-B));
v_func = matlabFunction(v, 'Vars', {z, t});
t_values = linspace(0.1, 6, 100);
max_v_values = NaN(size(t_values));
for idx = 1:length(t_values)
t_val = t_values(idx);
z_max = 0.6 * t_val;
if z_max <= 0
continue;
end
z_values = linspace(0.01, z_max - 0.01, 100);
current_vmax = -Inf;
for z_val = z_values
```

```
current_v = v_func(z_val, t_val);
if abs(imag(current_v)) > 1e-6
continue;
end
current_v = real(current_v);
if current_v > current_vmax
current_vmax = current_v;
end
end
if current_vmax ~= -Inf
max_v_values(idx) = current_vmax;
end
end
figure;
subplot(2,1,1)
plot(t_values, max_v_values, 'LineWidth', 1.5);
xlabel('t');
ylabel('V');
title('real part of V_p_r');
grid on;
```